\documentclass[12pt]{article}
\usepackage{amsmath,amsthm,amsfonts,amssymb,amscd}
\usepackage{graphics}
\usepackage[latin1]{inputenc}

\begin{document}

\newcommand{\po}{{\partial}}
\newcommand{\oc}{{\mathcal O}}

\centerline{\huge\bf A note on the stochastic nature of} 

\vskip .1in

\centerline{\huge\bf  Feynman quantum paths}

\thispagestyle{empty}

\vskip .3in

\centerline{{\sc Luiz C.L. Botelho}}

\centerline{Departamento de Matemática Aplicada,}

\centerline{Instituto de Matemática, Universidade Federal Fluminense,}

\centerline{Rua Mario Santos Braga}

\centerline{24220-140, Niterói, Rio de Janeiro, Brazil}

\centerline{e-mail: botelho.luiz@superig.com.br}

\vskip .5in

\begin{abstract}
We propose a Fresnel stochastic white noise framework to analyze the stochastic nature of the Feynman paths entering on the Feynman Path Integral expression for the Feynman \linebreak Propagator of a particle quantum mechanically moving under a time-independent potential.
\end{abstract}

\

\noindent
{\bf Key words:} Feynman Path Integrals; Nelson Stochastic Mechanics; Stochasticx Calculus.

\section{The Fresnel stochastic nature of Feynman quantum paths}

Let us start our note by considering as a basic object associated to a quantum particle of mass $m$ the following white noise functional Fresnel-Feynman path integral defined on an ensemble of closed, quantum trajectories related to a white noise process with correlation function depending on the particle classical mass $m$
\begin{equation}
I_m[j]=\int_{n(0)=0}^{n(T)=0} D^F[\eta(\sigma)] \exp\left(im\int_0^T \frac12[\eta(\sigma)]^2 d\sigma\right) \exp \left( i \int_0^T j(\sigma)\eta(\sigma) d\sigma \right) \tag{1}
\end{equation}

Here $\eta(\sigma)$ are the quantum white-noise Feynman closed trajectories on $R^N$ defined for the propagation time interval $\sigma\in[0,T]$ and satisfying, the vanishing and-point Dirichlet condition $\eta(0)=\eta(T)=0$.  $j(\sigma)$ denotes a fixed external real valued path source on the Schwartz test function space  $D(0,T)$.

It is worth to call the reader attention that the normalized white-noise external source Feynman path integral eq.(1) can be straightforwardly evaluated through a random Fourier series expansion for the random white noise trajectory $\eta(\sigma)$, $0\le\sigma\le T$ with the exact result
\begin{equation}
\frac{I_m[j]}{I_m[0]}=\exp \left\{ \frac i{2m} \int_0^T (j(\sigma))^2 d\sigma \right\}
\tag{2}
\end{equation}

In order to analyse the stochastic nature of the Feynman paths associated to the quantum system defined by a particle (of a fixed mentonion mass\ $m$) under the presence of a potencial $V(x)$ on $R^D$, we firstly consider the well-defined unique system classical trajectory connecting the spatial end-points $x_1$ and $x_2$ in the time interval $T$. Namely
\begin{equation}
\begin{aligned}
m \frac{d^2 x^{CL}}{d\sigma^2} (\sigma) &= (-\nabla V)(x^{CL}(\sigma)) \\
x^{CL}(0) &= x_1 \\
x^{CL}(T) &= x_2 \end{aligned} \tag{3}
\end{equation}

We now introduce what we call the effective  potential, through the Taylor expansion below defined
\begin{equation}
\begin{aligned}
\, & V^{eff}(x,[x^{CL}(\sigma)]) \\
&\quad \overset{\text{definition}}{\equiv} \,\, V(x^{CL}(\sigma)+\sqrt\hbar \, x) - V(x^{CL}(\sigma)) \\
&\qquad - \sqrt\hbar [(\nabla V)(x^{CL}(\sigma))]x. \end{aligned} \tag{4}
\end{equation}

Note that this effective   potential also depends functionally on the system's classical trajectory $\{x^{CL}(\sigma)\}$.

We now introduce the Feynman quantum trajectories of our system which are defined mathematically as those paths $x^q(\sigma)$, formally functionals of the Feynman white-noise path $n(\sigma)$, through the Hamilton-Jacobi equation for the quantum trajectory ([1]), where $E$ denotes the classical system total energy. Namelly
\begin{equation}
\frac1{2m} |\nabla W^{eff}(x,[x^{CL}])|^2+V^{eff}(x,[x^{CL}(\sigma)])=E,\tag{5}
\end{equation}
with the white-noise ODE's stochastic equation. A formal Sturm-Liouville stochastic problem
\begin{equation}
\begin{aligned}
\frac{dx^q(\sigma)}{d\sigma} &= \left( \frac1m (\nabla W^{eff})(x^q(\sigma),[x^{CL}])\right)+\eta(\sigma) \\
x^q(0) &= x^q(T)=0 \\
\eta(0) &= \eta(T)=0. \end{aligned}\tag{6}
\end{equation}

We claim that the full path defined below as quantum fluctuations around the classical path with ``size''\, of order $(\hbar)^{1/2}$
\begin{equation}
x(\sigma)=x^{CL}(\sigma)+\sqrt\hbar(x^q(\sigma))\tag{7}
\end{equation}
can be mathematically used to be the set of paths that enter in the Feynman path integral expression for the quantum mechanical propagator, and leading straighforwardly to the expected result that on the asymptotic semi-classical limit $\hbar\to0$, the leading contribution comes solely from the classical path.

To show these results, we firstly consider the formal object written in full below:
\begin{align*}
 G(x_1,x_2,T) &= \exp \left(\frac i\hbar S[x^{CL}(\sigma)] \right) \\
&\quad \times \Big\{ \int_{x^q(0)=0}^{x^q(T)=0} D^F[x^q(\sigma)] \Big[ \int_{\eta(0)=0}^{\eta(T)=0} D^F[\eta(\sigma)]\exp \Big( im \int_0^T \frac12(\eta(\sigma))^2 d\sigma \Big) \\
&\quad \times \det_F \Big[ \frac1 d{d\bar\sigma}-\frac1m \nabla_x \Big( \frac\delta{\delta x_q(\bar\sigma)} W^{eff}(x_q(\sigma), [x^{CL}(\sigma)]) \Big) \Big] \\
&\quad \times \delta^{(F)} \Big[ \frac{dx^q(\sigma)}{d\sigma} - \frac1m(\nabla W^{eff})(x^q(\sigma),[x^{CL}])-\eta(\sigma) \Big] \Big] \Big\}  \\
&\quad \times \exp (-iET) \exp( i[W^{eff}(0,[x^{CL}])-W^{eff}(T,[x^{CL}])]) \tag{8}
\end{align*}

We will formally show that it satisfies the system's quantum Feynman propagator equation for the time interval $T$
\begin{align*}
\frac\po{\po T} G(x_1,x_2,T) &= \Big( -\frac{\hbar^2}{2m} \Delta+V) G(x_1,x_2,T) \\
G(x_1,x_2,T)_{T\to0} &= \delta^{(N)} [x_1-x_2]. \tag{9}
\end{align*}

In order to arrive at such result, we firstly compute exactly the white-noise path integtral with the following result (written entirely on terms of the quantum path $x_q(\sigma)$)

\begin{align*}
G(x_1,x_2,T) &= f(E,T) \exp(\frac i{\hbar} S(x^{CL}(\sigma))) \\
&\quad \times \Big\{ \int_{x^q(0)=0}^{x^q(T)=0} D^F[x^q(\sigma)] \exp \Big( \frac{im}2 \int_0^T \Big[ \frac{dx^q}{d\sigma}-\frac1m ( \nabla W^{eff}(x^q(\sigma),[x^{CL}]))
\Big ]^2 d\sigma \Big ) \\
&\quad \times \text{def}_f \Big[ \frac d{d\bar\sigma} - \frac1m \nabla_x \Big( \frac\delta{\delta x_q(\bar\sigma)} W^{eff}(x_q(\sigma),[x^{CL}])\Big) \Big]  \Big\}. \tag{10}
\end{align*}

The action functional on the Feynman path integral thus posseses the following explicit functional form
\begin{align*}
\, & im \int_0^T \left( \frac12 \left[ \frac{dx^q(\sigma)}{d\sigma} - \frac 1m \nabla W^{eff}(x^q(\sigma),[x^{CL}]) \right]^2 \right)(\sigma) d\sigma \\
&\quad = im \int_0^T \left( \frac12 \left( \frac{dx^q(\sigma)}{d\sigma} \right)^2 \right)(\sigma)d\sigma \\
&\qquad + \frac{im}{2m^2} \int_0^T (\nabla W^{eff}(x^q(\sigma),[x^{CL}]))^2 d\sigma \\
&\qquad - i \int_0^T \left( \nabla W^{eff}(x^q(\sigma),[x^{CL}]) \cdot \frac{dx^q(\sigma)}{d\sigma} \right) d\sigma. \tag{11}
\end{align*}

By using the Itô's rule for derivatives we can evaluate the stochastic integral on last line of eq.(11)
\begin{align*}
&- i \left( \int_0^T \left( \nabla W^{eff}(x^q(\sigma),[x^{CL}]) \frac{dx^q(\sigma)}{d\sigma} \right) d\sigma \right) \\
&\quad =-i \Big( W^{eff}(x^q(T),[x^{CL}])-W^{eff}(x^q(\sigma),[x^{CL}])) \Big) \\
&+ \frac i2 \left( \int_0^T (\Delta W^{eff})(x^q(\sigma),[x^{CL}])d\sigma \right). \tag{12}
\end{align*}

Through eq.(5), we simpligy further eq.(11). Namely:
\begin{align*}
\, & \frac{im}{2m^2}\int_0^T (\nabla W^{eff}(x^q(\sigma),[x^{CL}]))^2 d\sigma \\
&= \left( \frac i{2m} \right) (2m) \left[ \int_)^T \left( E-V^{eff}(x^q(\sigma),[x^{CL}] \right)(\sigma) d\sigma \right] \\
&= + (iET) - i \left( \int_0^T V^{eff}(x^q(\sigma),[x^{CL}])(\sigma)d\sigma \right). \tag{13}
\end{align*}

By noting now that on the Itô's stochastic calculus One must use the operational rule $\theta(0)=+\frac12$, one has the following expression for the functional determinant
\begin{align*}
\, & \text{det}_f \left[ \frac d{d\bar\sigma}-\frac1m(\Delta W^{eff})(x_q(\sigma),[x^{CL}]) \right] \\
&= \exp \left( -\frac{im}{+2m} \int_0^T d\sigma(\Delta W^{eff})(x_q(\sigma),[x^{CL}])\right).\tag{14}
\end{align*}

It is worth call attention that this term canceals out exactly with similar term (however with inversew signal) on eq.(12). Note that if one has used from the beginning the stochastic Stratonovick calculus, with the prescription $\theta(0)=0$, eq.(14) would be one and the usual rule of integration for parts would be applied to eq.(12). We think that this an important result: the Feynman path integral appears to be insensitive to the kind of Stochastic Calculus used to define it. The point is just to be consistent with the stochastic prescription being used which usually reflects it self on the choose of the ``valve''\,  of the distribution $\theta(\sigma)$ on the point $\sigma=0$.

By grouping togheter all the above results on eq.(10), one has the partial outcome:
\begin{align*}
G(x_1,x_2,T) &= \exp\left( \frac i\hbar S[x_{CL}(\sigma)] \right) \\
&\quad \times \Big \{ \int_{x^q(0)=0}^{x^q(T)=0} D^F[x^q(\sigma)] \exp \Big[ i \int_0^T \Big( \frac 12 m \frac{dx^q}{d\sigma} \Big)^2 (\sigma) \\
& \qquad -i \int_0^T V^{eff}(x^q(\sigma),[x^{CL}])(\sigma) d\sigma \Big] \Big\}. \tag{15}
\end{align*}

We now observe that through the classical motion equation eq.(3) and the definition of the effective potential eq.(4) and the formal invariance of translation under classical trajectories of the Feynman path measure (see ref. [2]) one can re-write eq.(14) in the usual Feynman form
\begin{align*}
G(x_1,x_2,T) &= \int_{x(0)=x_1}^{x(T)=x_2} \\
&\quad \times \exp \left\{ \frac i\hbar S[x(\sigma)]\right\}, \tag{16}
\end{align*}
with the classical system action
\begin{align*}
S[x(\sigma)] &= \frac12 m \int_0^T \left( \frac{dx}{d\sigma} \right)^2(\sigma) d\sigma \\
&\quad - \int_0^T V(x(\sigma))d\sigma.\tag{17}
\end{align*}

The above heuristic-mathematical methods (from a rigorous mathematical point of view [3]) manipulations shows our claims.

\section{Conclusions}

As a general conclusion of our short note on the stochastic nature of the Feynman path integral, we stress that we have substituted the whole machinery of Feynman path integrals for the somewhat classical stochastic equation eq.(6) driven by (still mathematically formal) a Fresnel quantum white noise $n(\sigma)$. This result may be of practical use for Monte-Carlo sampling evaluations of observable since the numerical approximated solution of eq. the quantum Fresnel stochastic eq.(6) appears a less formadable task than evaluating the Feynman Path integral eq.(16) directly. These claims are result that eq.(5) is a first order system of usual non-linear partial differential equations and eq.(6) reduces to non-linear algebraic equations for the Fourier Coeficients of the expansion of the closed quantum trajectory in Fourier series. Note that 
$$
\eta(\sigma)=\sum\limits_{n=1}^\infty e_n \sin \left( \frac{2n\pi}T \sigma \right).
$$

As a last remark we call the reader attention that in the case of a time-dependent potential (a non-conservative classical system), the 
own quantum nature is not a straightforward concept (it is an open quantum system). If in this case we adopt the natural generalization of our Hamilton-Jacobi equation for time-dependent potentials
\begin{align*}
\, & \frac1{2m} \left| \nabla W^{eff}(x,t,[x^{CL}]) \right|^2 \\
&\quad + V^{eff}(x,t,[x^{CL}])= \frac\po{\po t} W^{eff}(x,t,[x^{CL}]) \tag{18}
\end{align*}
and the time-dependent Itô's rule for the integration by parts below
\begin{align*}
\, &- i \left( \int_0^T(\nabla W^{eff}(x^q(\sigma),[x^{CL}]))\cdot \frac{dx^q(\sigma)}{d\sigma} \right) d\sigma \\
&= -i \left( W^{eff}(x^q(T),T,[x^{CL}])-W^{eff}(x^q(0),0,[x^{CL}]) \right) \\
&\quad + \frac i2 \left( \int_0^T (\Delta W^{eff}) (x^q(\sigma),\sigma,[x^{CL}]) d\sigma \right) \\
&\quad + i \left( \int_0^T \left( \frac{\po W^{eff}}{dt} \right) (x^q(\sigma),\sigma,[x^{CL}]) d\sigma \right) \tag{19}
\end{align*}
one has an additional anomaly factor $I^{\text{anomaly}}$ on the path-integral result and fully depending on the function $W^{eff}(x,t,[x^{CL}])$; namely
\begin{equation}
I^{\text{anomaly}} = \exp \left[ 2i\int_0^T \left(\frac{\po W^{eff}}{\po t} \right) (x^q(\sigma),\sigma)d\sigma \right].\tag{20}
\end{equation}

That could signals that one should modify eq.(5) or point at that quantization of time-dependent mechanical point particle systems on the real $m$ of path integrals (Feynman propagators) need additional care or worse, time-dependent one particle quantum system are somewhat problematic in their quantum mechanical interpetration ([4]). Additional studies on this case will appears elsewhere.

\newpage

\noindent
{\bf REFERENCES}

\begin{itemize}

\item[{[1]}] Luiz C.L. Botelho, Modern Physics Letter B, vol. {\bf 14}, n$^{}o$ 3, 73--78, (2000).

\item[{[2]}] L.S. Schulman, Techniques and Applications of Path Integration, John Erley \& Sons (1981).

\item[{[3]}] Luiz C.L. Botelho, A note on Feynman-Kac path integral representations for scalar wave motions-Random Operators and Stochastic Equations, vol. {\bf 21}, 271--292, (2013), DOI:10.1515/rose-2013-0012.

\item[{[4]}] Luiz C.L. Botelho, Il Nuovo Cimento, vol. {\bf 117} B, 37--59, n${}^o$ 1, Gennaio 2002.

-- Luiz C.L. Botelho, Modern Physics Letters B, vol. {\bf 12}, n${}^o$ 14 \& 15, 569--573, (1998).

-- Luiz C.L. Botelho and E.P. Silva, Phys. Rev. 58E, 1141--1143, Jul. 1998.

-- Luiz C.L. Botelho, Modern Physics Letters 16B, 793--806, Sep. 2002.

\end{itemize}

\end{document}